\documentclass[preprint, twocolumn]{article}
\usepackage{graphicx}
\usepackage{amsmath}

\begin{document}

\title{Scalable Ellipsoidal Classification for Biparite Quantum States}

\author{David A. Herrera-Mart\'i \\
Universidad de Valencia \\
heda@postal.uv.es}

\maketitle

\begin{abstract}
\textsl{The Separability Problem is approached from the perspective of Ellipsoidal Classification. A Density Operator of dimension N can be represented as a vector in a real vector space of dimension $N^{2}- 1$, whose components are the projections of the matrix onto some selected basis. We suggest a method to test separability, based on successive optimization programs. First, we find the Minimum Volume Covering Ellipsoid that encloses a particular set of properly vectorized bipartite separable states, and then we compute the Euclidean distance of an arbitrary vectorized bipartite Density Operator to this ellipsoid. If the vectorized Density Operator falls inside the ellipsoid, it is regarded as separable, otherwise it will be taken as entangled. Our method is scalable and can be implemented straightforwardly in any desired dimension. Moreover, we show that it allows for detection of Bound Entangled States.}
\end{abstract}

\section{Introduction}

Entanglement is an essentially new resource for communication that lies at the very heart of Quantum Information Theory. Among several other exploits, it has been shown to permit information transfer at higher rates than classically expected, thank to the existence of quantum correlations between separated parties \cite{bennett1999eac,holevo2002eac,macchiavello2002eei}. Thus, a theory of Entanglement which offers a qualitative description as well as quantitative measures is highly desirable.

For pure states, a satisfactory theory exists and there are procedures both to detect and quantify the entanglement of a given state. For instance, the \emph{Von Neumann Entropy} of any of the reduced Density Operators gives a measure of the entanglement contained in the whole system, being zero if and only if the state is separable, and is directly related to the \emph{Schmidt Decomposition} of the Density Operator. For mixed states of dimension less or equal than 6, i.e. for $2\times 2$ and $2\times 3$ systems, \emph{Positivity under Partial Transposition} (PPT) is a necessary and sufficient condition for separability \cite{peres1996scd,horodecki1996sms}. However, in the general case, things are more involved, and it has been shown that, for dimension greater than 6, the problem of checking whether a mixed state is entangled or separable is NP-hard \cite{gurvits2002qmt}. Thus, one should not expect to find an exact solution to the Separability Problem.

In the recent years some approximations based on different relaxations of the problem have been suggested. Doherty \emph{et al.}\cite{doherty2002dsa,doherty2004cfs} proposed a test based on nested semidefinite programs to search for symmetric extensions of a given bipartite system, which is a stronger condition for separability than the PPT criterion. Brand\~{a}o and Vianna showed that the Separability Problem can be cast as a robust semidefinite program, which is an NP-hard problem, and proposed both probabilistic and deterministic relaxations \cite{branduao2004rsp,branduao2004smm}. Note that these tests only provide bounds to the certainty of the answer to the question  ``is the state $\rho$ separable or entangled?", but they don't offer a quantification of the entanglement contained in $\rho$.

Our method relies on the concept of the \emph{Minimum Volume Covering Ellipsoid} (also known as the L\"{o}wner-John Ellipsoid) of a given set of vectors in vector space $\mathbf{R}^{N^{2}-1}$, isomorphic to the space of Density Operators of dimension N. We show that, for vectors representing separable bipartite Density Operators, it is possible to identify this ellipsoid with the boundary of the set of bipartite separable states, up to some accuracy. Interestingly, this approach can be used to estimate the amount of entanglement contained in $\rho$.

In next section, we briefly review fundamental issues concerning geometric entanglement measures. In section 3 we introduce our method and apply it to $2\times 2$ and $2\times 3$ systems. In section 4, we show that it allows for detection of Bound Entanglement.

\section{Review of Geometric Entanglement Measures}

The study of entanglement from the geometrical point of view offers insightful ways both to detect and to quantify quantum correlations between different systems. First one is known as \emph{Entanglement Witnesses} (EW), which split the space of states in two parts in such a way that the set of separable states lies strictly inside one of the resulting half-spaces. We denote the set of all Density Operators by $\mathcal{D}$, and the set of separable states by $\mathcal{S}$. It follows from probability theory that both sets are convex \cite{holevo1985pas}. Let $W=W^{\dagger}$ be an EW, then:

\begin{equation}\label{rho}
    \langle W, \rho   \rangle <    0 \text{, for some $\rho \in \mathcal{D}\setminus\mathcal{S}$}
\end{equation}

\begin{equation}\label{sigma}
    \langle W, \sigma \rangle \geq 0 \text{, $\forall  \sigma \in \mathcal{S}$}
\end{equation}

Clearly this defines a hyperplane dividing $\mathcal{D}$: $\langle W, \sigma \rangle \geq \langle W, \rho \rangle$. Here $\langle.\rangle$ is the inner product for matrices $\langle A, B \rangle = \text{Tr}(A^{\dagger}B)$. This product naturally defines the Frobenius norm $\|A\|_{F} = \sqrt{\text{Tr}(A^{\dagger}A)}$

Another measure is the \emph{Hilbert-Schmidt Distance}, $D_{HS}$, of a Density Operator to the set of separable states $\mathcal{S}$:

\begin{equation}\label{distance}
D_{HS} = \min_{\sigma \in \mathcal{S}} \|\rho - \sigma\|_{F}
\end{equation}

Note that computing $D_{HS}$ involves the obtention of $\sigma_{0}=P_{\mathcal{S}}(\rho)$, the projection of $\rho$ onto $\mathcal{S}$. An Optimal Entanglement Witness, that is, an EW for which (\ref{rho}) holds for the largest number of states in $\mathcal{D}\setminus\mathcal{S}$, will be tangent to $\mathcal{S}$ at this point.

\subsection{Projection of Density Operators onto a Vector Space}

Any Density Operator $\rho$ of dimension N can be written in vectorized form:

$$\rho = \frac{1}{N}(\mathbf{1}_{N} + \gamma \sum^{N^{2}- 1}_{i=1} r_{i} \sigma_{i})$$

where $\mathbf{1}_{N}$ is the unit matrix of dimension N and $\{\sigma_{i}\}^{N^{2}- 1}_{i}$ is some basis of self-adjoint, trace-free matrices, such that:

\begin{equation}\label{trace}
    \text{Tr}(\sigma_{i}\sigma_ {j}) = \alpha \delta_{ij}
\end{equation}

\begin{equation}\label{component}
    r_{i}=\frac{N}{\alpha} \text{Tr}(\sigma_ {i} \rho)
\end{equation}

and $\gamma$ is chosen to normalize $\mathbf{r}$. The Frobenius norm induces an Euclidean norm in $\mathbf{R}^{N^{2}- 1}$. For pure states, $ \| \mathbf{r}\|_{2} = 1$, while for mixed states $ \| \mathbf{r}\|_{2} < 1$. This illustrates the convexity of the set of states.

Thus, a one-to-one relation between Density Operators of dimension N and vectors in $\mathbf{R}^{N^{2}- 1}$ can been established. This allows us to switch from the matrix space picture to the real vector space picture, which will prove useful for our method.

\subsection{Duality between Hilbert-Schmidt Distance and Entanglement Witnesses}

It is a general result from geometric optimization that the problem of finding a separating hyperplane between a point $\mathbf{p}$ and  a convex set $\mathcal{C}$ is dual to the problem of finding the distance between $\mathcal{C}$ and $\mathbf{p}$ \cite{boyd2004co}. In matrix space language, this duality can be illustrated as follows. Let $\rho$ and $\mathcal{S}$ be an arbitrary Density Operator and the set of separable states. The problem (\ref{distance}) can be posed as:

\begin{eqnarray}\label{distance2}
    \min &\| \tau \|_{F}\\
    \text{such that}&\rho - \sigma =\tau \nonumber \\
    &\sigma \in \mathcal{S}\nonumber
\end{eqnarray}

with variables $\tau$ and $\sigma$. The Lagrangian of (\ref{distance2}) is:

\begin{equation}\label{lagrangian}
    \mathcal{L} = \| \tau \|_{F} + \langle W,\rho - \sigma - \tau\rangle
\end{equation}

where W is the Lagrange multiplicator associated to the equality constraint. It is not hard to see that it represents a a hyperplane. Noting that $\langle W,\tau \rangle \leq \| W \|_{F}\| \tau \|_{F}$, the dual function can be written as:

\begin{equation}\label{dualfunction}
    g(W) = \min_{\sigma \in \mathcal{S},\tau} [\langle W,\rho \rangle - \langle W,\sigma \rangle + \| \tau \|_{F}(1-\| W \|_{F} + \delta)]
\end{equation}

where the parameter $\delta\geq 0$ is related to the relative orientation between the hyperplane represented by W, and the line going from the separable set $\mathcal{S}$ to the Density Operator $\rho$, and it is equal to zero if and only if they are perpendicular. For (\ref{dualfunction}) to be bounded from below in $\tau$ we must include the additional constraint $\| W \|_{F} - \delta \leq 1$ . So the dual problem of (\ref{distance2}) is:

\begin{equation}\label{dualproblem}
  \max_{\| W \| - \delta \leq 1}[ \min_{\sigma \in \mathcal{S}} [\langle W,\rho \rangle - \langle W,\sigma \rangle]]
\end{equation}

It is straightforward to check that the optimal value of (\ref{dualproblem}) is attained if and only if W is an Optimal EW. This result, also known as the Bertlmann-Narnhoffer-Thirring Theorem (see Ref. \cite{bertlmann2002gpe}), will let us trace a link between the entanglement detection and quantification problems (compare also with Refs. \cite{brandao2005qew,eisert2007qew})

\section{Ellipsoidal Classification}

The basic premise of our method is that the set of separable states $\mathcal{S}$ can be approximated by a \emph{Minimum Volume Covering Ellipsoid} (MVCE) of an ensemble of vectors corresponding to some separable Density Operators. Then, the following classification scheme can be adopted: if a vector falls inside the MVCE, it will be taken as separable, and if it falls outside, it will be regarded as entangled.

An ellipsoid centered at $\mathbf{x}_{c}$ can be expressed as:

\begin{equation}\label{ellipsoid}
    \mathcal{E}= \{ \mathbf{x} | (\mathbf{x} - \mathbf{x}_{c})^{T}A(\mathbf{x} - \mathbf{x}_{c}) \leq 1 \}
\end{equation}

where $A = A^{T}$ is a positive definite matrix of dimension $N^{2}- 1$. The volume of this ellipsoid is proportional to $\det(A^{-1/2})$.

Since in matrix space quadratic forms are not defined, we need to work in a real vector space to build this ellipsoid. We first obtain an ensemble of ``separable vectors" by means of tensorially multiplying states along all directions specified by some canonical basis. For instance, in the $2\times 2$ case, this ensemble would be $\{\mathbf{x}^{sep}_{i}\} =\{(1,0,0)\otimes(1,0,0), (1,0,0)\otimes(-1,0,0),(1,0,0)\otimes(0,1,0), (1,0,0)\otimes(0,-1,0), ...,(0,0,-1)\otimes(0,0,-1)\}$ . Later on we will see that it is convenient to vary the norm $\|\mathbf{x}^{sep}_{i}\|_{2}$ of these vectors. This procedure ensures that all vectors will lie as spaced as possible in the separable set $\mathcal{S}$. Secondly, we minimize the volume of an ellipsoid, constrained to have all generated ``separable vectors" falling inside it. One way to obtain the MVCE of this ensemble would be to solve the following problem:

\begin{eqnarray}\label{mvce}
    \min &\log\det A^{-1}\\
    \text{such that}&(\mathbf{x}^{sep}_{i}-\mathbf{x}_{c})^{T}A(\mathbf{x}^{sep}_{i}-\mathbf{x}_{c})\leq 1 \nonumber
\end{eqnarray}

with variables A and $\mathbf{x}_{c}$. Here, logarithm was taken in order to drop off proportionality terms. Despite the exponential growth of the dimension of the associated vector space, interior point methods used for minimization still converge polynomially to a solution in dimension as large as 1000, or more \cite{boyd2004co}. Throughout this work, we used the Disciplined Convex Programming Rule-Set implemented in the Matlab Package CVX \cite{cvx}.

\begin{figure}
  \includegraphics[width=7.5cm]{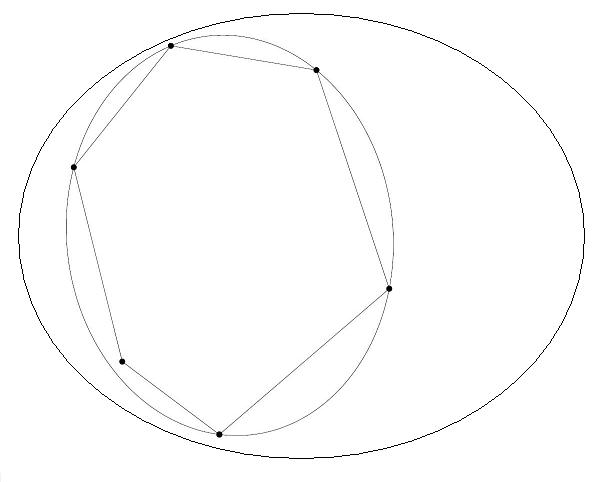}\\
  \caption{The vertices of the polytope are the generated ``separable vectors" of which we find the MVCE. The larger set corresponds to the whole space of Density Operators}\label{DOspace}
\end{figure}

\subsection{Results for $2\times 2$ and $2\times 3$ Systems}

The Separability Problem is solved for $2\times 2$ and $2\times 3$ systems, thanks to the PPT criterion. We use this fact to benchmark our method. For a sample of 1000 ``separable vectors" and 1000 ``entangled vectors", we solved two different problems: first one is the original problem (\ref{distance}) casted as:

\begin{eqnarray}\label{distance3}
    \min &\|\rho - \sigma\|_{F} \\
    \text{such that} &\sigma^{T_{A}}\geq 0 \nonumber
\end{eqnarray}

which gives the true results. The variable is $\sigma$, and $\sigma^{T_{A}}$ denotes Partial Transpose in subsystem A. The second problem was to find the MVCE through (\ref{mvce}), and compute the distance to this ellipsoid in a similar way:

\begin{eqnarray}\label{distance4}
  \min &\|\mathbf{r} - \mathbf{s}\|_{2} \\
  \text{such that} &(\mathbf{s}-\mathbf{x}_{c})^{T}A(\mathbf{s}-\mathbf{x}_{c})\leq 1 \nonumber
\end{eqnarray}

where $\mathbf{r}$ and $\mathbf{s}$ stand for the vectorized counterparts of $\rho$ and $\sigma$. The results obtained for pure vectors ($\|\mathbf{x}^{sep}_{i}\|_{2} = 1 $) are rather discouraging: whereas none of the generated ``separable vectors" fell outside the MVCE, only 12.7\% of the ``entangled vectors" were detected. This led us to shrink our ellipsoid by reducing the norm of the generated ensemble $\{\mathbf{x}^{sep}_{i}\}$. At the expense of letting some ``separable vectors" fall outside the ellipsoid, we increase the number of correctly classified ``entangled vectors". This recalls a \emph{binary hypothesis test}, and we shall adopt the corresponding terminology to expose our results. The event that a true ``separable vector" falls outside the MVCE will be a \emph{false positive}, while if an ``entangled vector" falls inside the  MVCE, it will be \emph{false negative}. Stepwise reducing the norm of the vectors belonging to the separable ensemble we obtained Tables 1 and 2.

\begin{table}
  \centering
\begin{tabular}{|c||c|c|}
  \hline
  \multicolumn{3}{|c|}{2 x 2 Systems}\\
  \hline
  Norm & False Positives & False Negatives \\
  \hline \hline
  0.1 & 962 & 0 \\
  0.2 & 868 & 0 \\
  0.3 & 687 & 0 \\
  0.4 & 484 & 0 \\
  0.5 & 287 & 15 \\
  0.6 & 180 & 184 \\
  0.7 & 92 & 410 \\
  0.8 & 32 & 600 \\
  0.9 & 5 & 755 \\
  1.0 & 0 & 873 \\
  \hline
\end{tabular}
\caption{Number of misclassified vectors in a sample of 1000 ``separable vectors" and 1000 ``entangled vectors", as a function of the Euclidean norm of the vectors of the separable ensemble}
\end{table}

\begin{table}
  \centering
\begin{tabular}{|c||c|c|}
  \hline
  \multicolumn{3}{|c|}{2 x 3 Systems}\\
  \hline
  Norm & False Positives & False Negatives \\
  \hline \hline
  0.1 & 949 & 0 \\
  0.2 & 812 & 0 \\
  0.3 & 597 & 0 \\
  0.4 & 427 & 52 \\
  0.5 & 269 & 196 \\
  0.6 & 160 & 389 \\
  0.7 & 80 &  572 \\
  0.8 & 34 & 699 \\
  0.9 & 11 & 807 \\
  1.0 & 0 & 900 \\
  \hline
\end{tabular}
\caption{Number of misclassified vectors in a sample of 1000 ``separable vectors" and 1000 ``entangled vectors", as a function of the Euclidean norm of the vectors of the separable ensemble}
\end{table}

\begin{figure}
  \includegraphics[width=9cm]{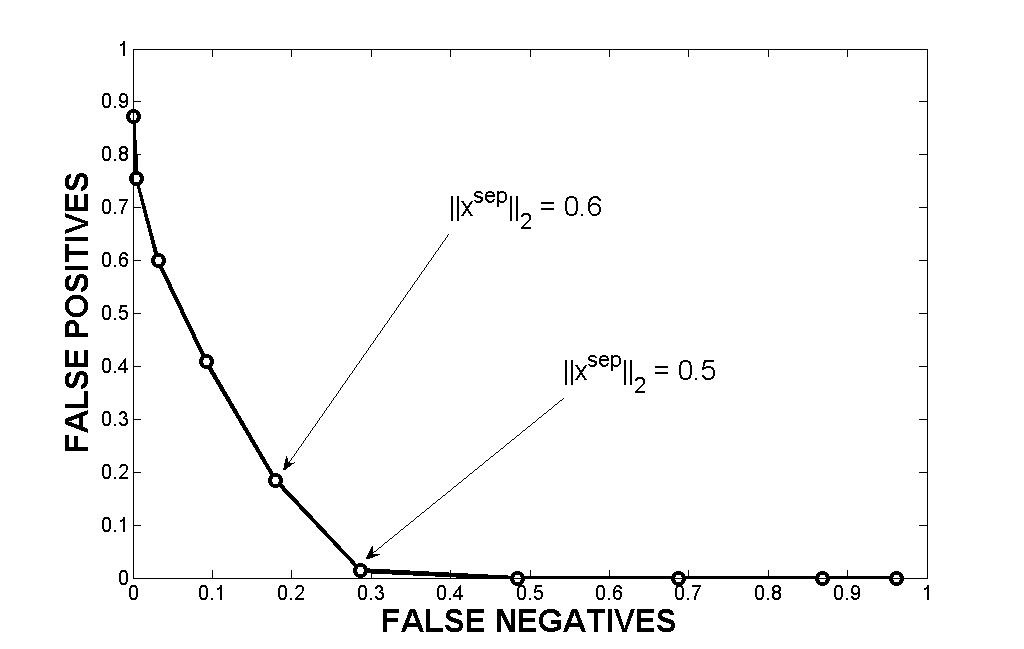}\\
  \caption{False Negatives versus False Positives for $2 \times 2$ systems, showing that there exists an area where the probability of wrongly classifying a vector can be brought down to 15.1\%, between $\|\mathbf{x}^{sep}_{i}\|_{2} = 0.6$ and $\|\mathbf{x}^{sep}_{i}\|_{2} = 0.5$}\label{fnfp22}
\end{figure}

\begin{figure}
  \includegraphics[width=9cm]{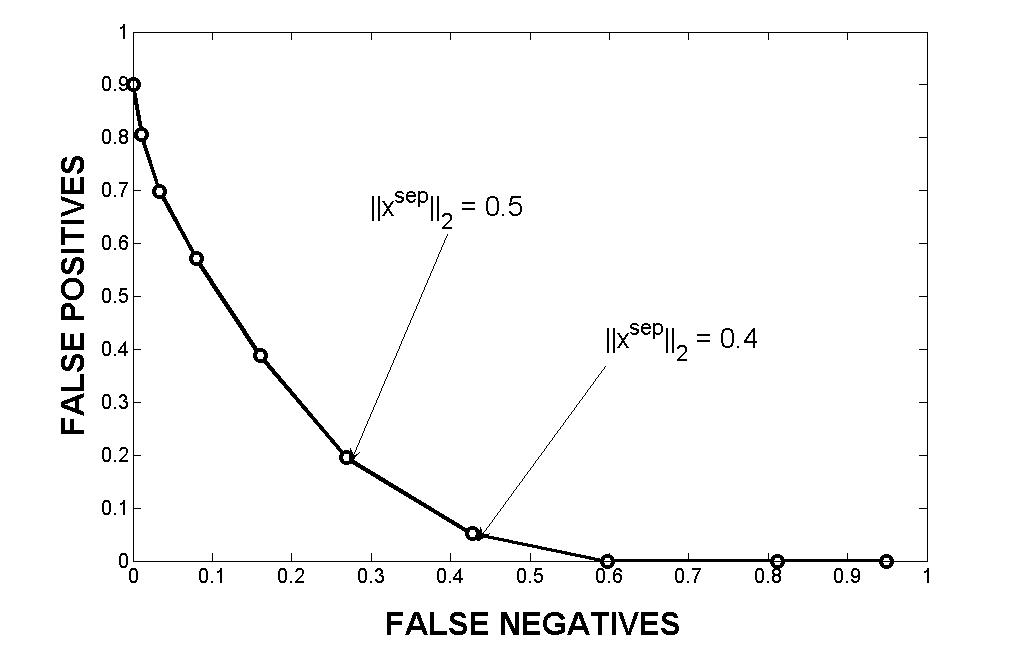}\\
  \caption{False Negatives versus False Positives for $2 \times 3$ systems. The error probability can be reduced to 23.2\%, between $\|\mathbf{x}^{sep}_{i}\|_{2} = 0.5$ and $\|\mathbf{x}^{sep}_{i}\|_{2} = 0.4$}\label{fnfp23}
\end{figure}

There is a trade-off between the number of correctly classified states and non-ambiguousness of the test. The relevant area of $2\times 2$ systems is between norms  $\|\mathbf{x}^{sep}_{i}\|_{2} = 0.6$ and $\|\mathbf{x}^{sep}_{i}\|_{2} = 0.5$, as can be seen in Fig. \ref{fnfp22}. A measure of entanglement should be as unambiguous as possible, and thus we argue that the best choice is $\|\mathbf{x}^{sep}_{i}\|_{2} = 0.5$, since for this case only about 1.5\% of the ``entangled vectors" are misclassified. For this choice, in general, a vector will be misclassified 15.1\% of the time. For $2\times 3$ systems (see Fig. \ref{fnfp23}), the MCVE approximates somewhat less efficiently the separable set. However, still 76.8\% of the vectors are correctly classified, in the area comprised between $\|\mathbf{x}^{sep}_{i}\|_{2} = 0.5$ and $\|\mathbf{x}^{sep}_{i}\|_{2} = 0.4$. In these systems, it misclassifies at least 5.2\% of the ``separable vectors".

\subsection{Pseudo-Entanglement Witnesses}

For a vector space endowed with the Euclidean norm, there is a simple way to construct a tangent hyperplane to a given ellipsoid. We can use this fact to build realistic observables amenable to a laboratory setting.

The tangent hyperplane to the ellipsoid can be expressed as:

\begin{equation}\label{tangent1}
    \nabla_{\mathbf{x}}[(\mathbf{x}-\mathbf{x}_{c})^{T}A(\mathbf{x}-\mathbf{x}_{c}) - 1]_{s_{0}}(\mathbf{r} - \mathbf{s}_{0}) = 0
\end{equation}

where $\mathbf{s}_{0} = P_{\mathcal{E}}(\mathbf{r})$ is the projection of the vectorized Density Operator under study onto the MVCE. It can be expressed in affine form as:

\begin{equation}\label{tangent2}
    (\mathbf{s}_{0} - \mathbf{x}_{c})^{T}{A(\mathbf{r} - \mathbf{x}_{c}) = 1}
\end{equation}

\begin{figure}
  \includegraphics[width=9cm]{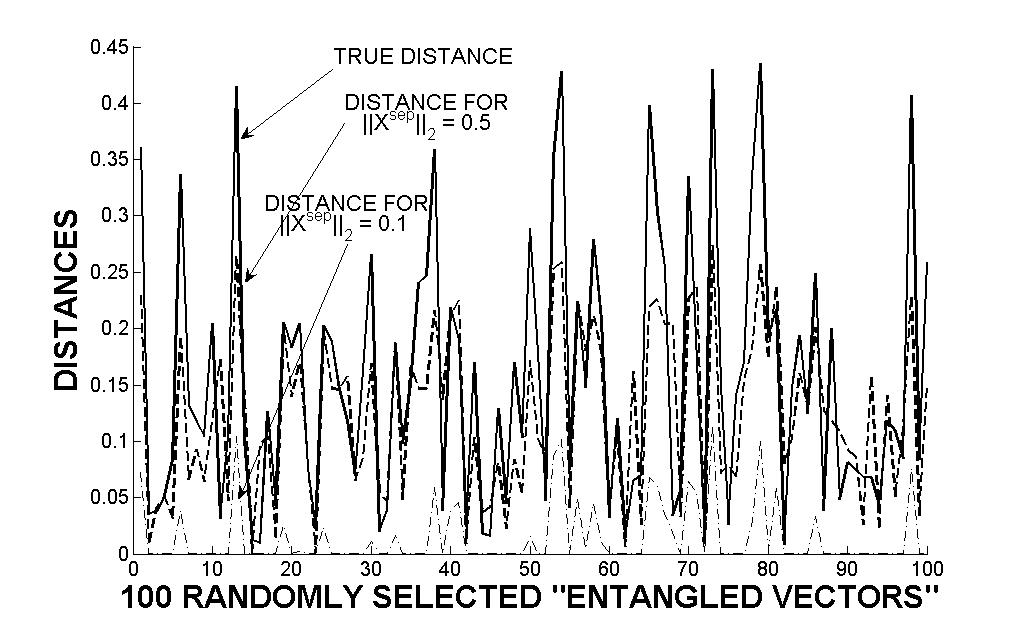}\\
  \caption{For 100 random ``entangled vectors" of a $2 \times 2$ system, the continuous black line is the true distance to the separable set $\mathcal{S}$, while the dashed line stands for the distance of the vectors for a MVCE of ``separable vectors" of norm $\|\mathbf{x}^{sep}_{i}\|_{2} = 0.5$. At the bottom, the pointed line represents the distances obtained for norm $\|\mathbf{x}^{sep}_{i}\|_{2} = 1$}\label{bounds}
\end{figure}

(compare with Ref. \cite{bertlmann2005oew}). It is important to keep in mind that, although the hyperplanes introduced in (\ref{tangent2}) very much resemble an Entanglement Witness, they are not so in general. This is because our MCVE may in general be a proper subset of the separable set $\mathcal{S}$, and no tangent hyperplane to this MVCE will strictly separate $\mathcal{S}$ from any entangled state. Nevertheless, these \emph{Pseudo}-EW can be used to estimate the amount of entanglement contained in a given entangled matrix $\rho$ via (\ref{dualproblem}), which at the optimal value will be equal to (\ref{distance2})\cite{boyd2004co}. For an illustration of entanglement estimation see Fig. \ref{bounds}.

\section{Bound Entanglement Detection}

For composite systems of dimension higher than 6, there is a special kind of entangled states that cannot be used, in principle, to enhance communication. The entanglement contained in these states cannot be distilled to obtain pure entangled states \cite{horodecki1998mse}, and it receives the name of \emph{Bound Entanglement} (BE). The PPT criterion fails to detect this kind of states, and it just becomes a necessary condition for quantum correlations to arise. Other criteria to detect and quantify BE have been proposed, such as non-decomposable EW \cite{lewenstein2000oew}, Schmidt number Witnesses \cite{sanpera2001snw}, and, more recently, a geometric approach based on separating hyperplanes \cite{bertlmann2008gew}.

Our approach is in the spirit of the latter of the aforementioned methods, but instead of hyperplanes, we shall use the MVCE in order to detect BE. Intuitively, the ellipsoid covering a set of ``separable vectors" should leave bound entangled states on its outside. This fact is studied in 3X3 systems, where a parametrization of bound entangled states, due to P. Horodecki, is available \cite{horodecki1997sca}. These states $\rho_{BE}$ depend on a scalar $a \in [0,1]$, and are given by:

$$\rho_{BE}(a) = \frac{1}{8a + 1}
\left(
  \begin{array}{ccccccccc}
    a & 0 & 0 & 0 & a & 0 & 0 & 0 & a \\
    0 & a & 0 & 0 & 0 & 0 & 0 & 0 & 0 \\
    0 & 0 & a & 0 & 0 & 0 & 0 & 0 & 0 \\
    0 & 0 & 0 & a & 0 & 0 & 0 & 0 & 0 \\
    a & 0 & 0 & 0 & a & 0 & 0 & 0 & a \\
    0 & 0 & 0 & 0 & 0 & a & 0 & 0 & 0 \\
    0 & 0 & 0 & 0 & 0 & 0 & \frac{1 + a}{2} & 0 & \frac{\sqrt{1 - a^{2}}}{2} \\
    0 & 0 & 0 & 0 & 0 & 0 & 0 & a & 0 \\
    a & 0 & 0 & 0 & a & 0 & \frac{\sqrt{1 - a^{2}}}{2} & 0 & \frac{1 + a}{2} \\
  \end{array}
\right)
$$

Surprisingly, for norms of the generated separable ensemble of 0.6 and below, all bound entangled states are detected. The obtained results are shown in Table 3.

\begin{table}
  \centering
\begin{tabular}{|c||c|}
  \hline
  \multicolumn{2}{|c|}{3 x 3 Systems}\\
  \hline
  Norm &  Detected States\\
  \hline \hline
  0.1 & 1000\\
  0.2 & 1000\\
  0.3 & 1000\\
  0.4 & 1000\\
  0.5 & 1000\\
  0.6 & 1000\\
  0.7 & 226\\
  0.8 & 149\\
  0.9 & 107\\
  1.0 & 79\\
  \hline
\end{tabular}
\caption{1000 bound entangled states were generated, with parameter ``a" running from 0.001 to 1. The distances of the associated vectors to the different MVCEs were obtained. For norms of the separable ensemble of 0.6 and below, all bound entangled states were detected}
\end{table}

\begin{figure}
  \includegraphics[width=9cm]{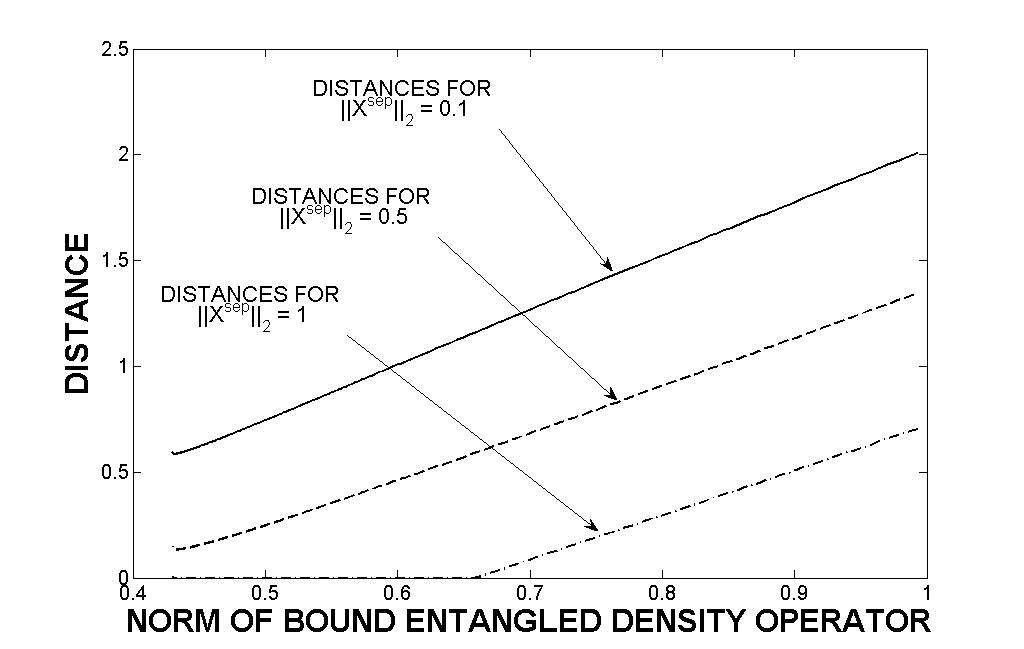}\\
  \caption{There is a linear dependence between the distance to the MVCE and the norm of the associated Density Operator}\label{BEdist}
\end{figure}

As expected, the distance to the MVCE of the detected states linearly depends on the norm of the associated Density Operator. This interdependence is depicted in Fig. \ref{BEdist}

Although, for sufficiently small norms, all BE states are detected, one should keep in mind that, in these cases, an unacceptable number of false positives is introduced. This fact precludes a truly reliable criterion. On top of this, not all types of BE states were tested, so no complete evidence for detection of BE is provided. Nevertheless, this gives a quantitative picture to the intuitive idea that, since BE states lie outside the separable set, some of them will necessarily be detected by our MVCE method.

\section{Conclusions and Further Perspectives}

Our method presents some drawbacks. First important one is ambiguousness, since false positives must be necessarily introduced, to some extent. Also, the efficiency of the MVCE decreases as $1/N$ as the dimensionality of the problem grows. This can be compensated reducing the norm of the ``separable vectors" belonging to the generated ensemble. Although the reduction of the norm is geometrically justified, we are still lacking a qualitative physical argumentation, if there is any.

On the other hand, it presents at least two appealing advantages. Interior point methods behave well in very high dimensions, so in principle any discrete quantum system can be accounted for. For this reason the method is \emph{scalable}. Moreover, relatively few resources are needed in contrast to other methods, and in fact, once the MVCE is obtained, a decision is made in very few seconds.

This method could probably be refined allowing some robust quadratic classification, or using a polyhedral approximation instead of computing the MVCE. However, this has the taste of being NP-hard.

The author acknowledges fruitful conversations with B. Beferull, F. Brand\~{a}o and J. de Vicente.

\bibliographystyle{numbersdadqw}

\begin{thebibliography}{22}
\expandafter\ifx\csname natexlab\endcsname\relax\def\natexlab#1{#1}\fi
\expandafter\ifx\csname bibnamefont\endcsname\relax
  \def\bibnamefont#1{#1}\fi
\expandafter\ifx\csname bibfnamefont\endcsname\relax
  \def\bibfnamefont#1{#1}\fi
\expandafter\ifx\csname citenamefont\endcsname\relax
  \def\citenamefont#1{#1}\fi
\expandafter\ifx\csname url\endcsname\relax
  \def\url#1{\texttt{#1}}\fi
\expandafter\ifx\csname urlprefix\endcsname\relax\def\urlprefix{URL }\fi
\providecommand{\bibinfo}[2]{#2}
\providecommand{\eprint}[2][]{\url{#2}}

\bibitem[{1}]{bennett1999eac}
\bibinfo{author}{\bibfnamefont{C.}~\bibnamefont{Bennett}},
  \bibinfo{author}{\bibfnamefont{P.}~\bibnamefont{Shor}},
  \bibinfo{author}{\bibfnamefont{J.}~\bibnamefont{Smolin}}, \bibnamefont{and}
  \bibinfo{author}{\bibfnamefont{A.}~\bibnamefont{Thapliyal}},
  \bibinfo{journal}{Physical Review Letters} \textbf{\bibinfo{volume}{83}},
  \bibinfo{pages}{3081} (\bibinfo{year}{1999}).

\bibitem[{2}]{holevo2002eac}
\bibinfo{author}{\bibfnamefont{A.}~\bibnamefont{Holevo}},
  \bibinfo{journal}{Journal of Mathematical Physics}
  \textbf{\bibinfo{volume}{43}}, \bibinfo{pages}{4326} (\bibinfo{year}{2002}).

\bibitem[{3}]{macchiavello2002eei}
\bibinfo{author}{\bibfnamefont{C.}~\bibnamefont{Macchiavello}}
  \bibnamefont{and} \bibinfo{author}{\bibfnamefont{G.}~\bibnamefont{Palma}},
  \bibinfo{journal}{Physical Review A} \textbf{\bibinfo{volume}{65}},
  \bibinfo{pages}{50301} (\bibinfo{year}{2002}).

\bibitem[{4}]{peres1996scd}
\bibinfo{author}{\bibfnamefont{A.}~\bibnamefont{Peres}},
  \bibinfo{journal}{Physical Review Letters} \textbf{\bibinfo{volume}{77}},
  \bibinfo{pages}{1413} (\bibinfo{year}{1996}).

\bibitem[{5}]{horodecki1996sms}
\bibinfo{author}{\bibfnamefont{M.}~\bibnamefont{Horodecki}},
  \bibinfo{author}{\bibfnamefont{P.}~\bibnamefont{Horodecki}},
  \bibnamefont{and}
  \bibinfo{author}{\bibfnamefont{R.}~\bibnamefont{Horodecki}},
  \bibinfo{journal}{Physics Letters A} \textbf{\bibinfo{volume}{223}},
  \bibinfo{pages}{1} (\bibinfo{year}{1996}).

\bibitem[{6}]{gurvits2002qmt}
\bibinfo{author}{\bibfnamefont{L.}~\bibnamefont{Gurvits}},
  \bibinfo{journal}{Arxiv preprint quant-ph/0201022}  (\bibinfo{year}{2002}).

\bibitem[{7}]{doherty2002dsa}
\bibinfo{author}{\bibfnamefont{A.}~\bibnamefont{Doherty}},
  \bibinfo{author}{\bibfnamefont{P.}~\bibnamefont{Parrilo}}, \bibnamefont{and}
  \bibinfo{author}{\bibfnamefont{F.}~\bibnamefont{Spedalieri}},
  \bibinfo{journal}{Physical Review Letters} \textbf{\bibinfo{volume}{88}},
  \bibinfo{pages}{187904} (\bibinfo{year}{2002}).

\bibitem[{8}]{doherty2004cfs}
\bibinfo{author}{\bibfnamefont{A.}~\bibnamefont{Doherty}},
  \bibinfo{author}{\bibfnamefont{P.}~\bibnamefont{Parrilo}}, \bibnamefont{and}
  \bibinfo{author}{\bibfnamefont{F.}~\bibnamefont{Spedalieri}},
  \bibinfo{journal}{Physical Review A} \textbf{\bibinfo{volume}{69}},
  \bibinfo{pages}{22308} (\bibinfo{year}{2004}).

\bibitem[{9}]{branduao2004rsp}
\bibinfo{author}{\bibfnamefont{F.}~\bibnamefont{Brand\~{a}o}} \bibnamefont{and}
  \bibinfo{author}{\bibfnamefont{R.}~\bibnamefont{Vianna}},
  \bibinfo{journal}{Physical Review A} \textbf{\bibinfo{volume}{70}},
  \bibinfo{pages}{62309} (\bibinfo{year}{2004}{\natexlab{a}}).

\bibitem[{10}]{branduao2004smm}
\bibinfo{author}{\bibfnamefont{F.}~\bibnamefont{Brand\~{a}o}} \bibnamefont{and}
  \bibinfo{author}{\bibfnamefont{R.}~\bibnamefont{Vianna}},
  \bibinfo{journal}{Physical Review Letters} \textbf{\bibinfo{volume}{93}},
  \bibinfo{pages}{220503} (\bibinfo{year}{2004}{\natexlab{b}}).

\bibitem[{11}]{holevo1985pas}
\bibinfo{author}{\bibfnamefont{A.}~\bibnamefont{Holevo}},
  \emph{\bibinfo{title}{{Probabilistic and Statistical Aspects of Quantum
  Theory}}} (\bibinfo{publisher}{North-Holland Publishing Company},
  \bibinfo{year}{1985}).

\bibitem[{12}]{boyd2004co}
\bibinfo{author}{\bibfnamefont{S.}~\bibnamefont{Boyd}} \bibnamefont{and}
  \bibinfo{author}{\bibfnamefont{L.}~\bibnamefont{Vandenberghe}},
  \emph{\bibinfo{title}{{Convex Optimization}}} (\bibinfo{publisher}{Cambridge
  University Press}, \bibinfo{year}{2004}).

\bibitem[{13}]{bertlmann2002gpe}
\bibinfo{author}{\bibfnamefont{R.}~\bibnamefont{Bertlmann}},
  \bibinfo{author}{\bibfnamefont{H.}~\bibnamefont{Narnhofer}},
  \bibnamefont{and} \bibinfo{author}{\bibfnamefont{W.}~\bibnamefont{Thirring}},
  \bibinfo{journal}{Physical Review A} \textbf{\bibinfo{volume}{66}},
  \bibinfo{pages}{32319} (\bibinfo{year}{2002}).

\bibitem[{14}]{brandao2005qew}
\bibinfo{author}{\bibfnamefont{F.}~\bibnamefont{Brand\~{a}o}},
  \bibinfo{journal}{Physical Review A} \textbf{\bibinfo{volume}{72}},
  \bibinfo{pages}{22310} (\bibinfo{year}{2005}).

\bibitem[{15}]{eisert2007qew}
\bibinfo{author}{\bibfnamefont{J.}~\bibnamefont{Eisert}},
  \bibinfo{author}{\bibfnamefont{F.}~\bibnamefont{Brand\~{a}o}},
  \bibnamefont{and}
  \bibinfo{author}{\bibfnamefont{K.}~\bibnamefont{Audenaert}},
  \bibinfo{journal}{New Journal of Physics} \textbf{\bibinfo{volume}{9}},
  \bibinfo{pages}{46} (\bibinfo{year}{2007}).

\bibitem[{16}]{cvx}
\bibinfo{author}{\bibfnamefont{M.}~\bibnamefont{Grant}} \bibnamefont{and}
  \bibinfo{author}{\bibfnamefont{S.}~\bibnamefont{Boyd}},
  \bibinfo{note}{http://stanford.edu/boyd/cvx}.

\bibitem[{17}]{bertlmann2005oew}
\bibinfo{author}{\bibfnamefont{R.}~\bibnamefont{Bertlmann}},
  \bibinfo{author}{\bibfnamefont{K.}~\bibnamefont{Durstberger}},
  \bibinfo{author}{\bibfnamefont{B.}~\bibnamefont{Hiesmayr}}, \bibnamefont{and}
  \bibinfo{author}{\bibfnamefont{P.}~\bibnamefont{Krammer}},
  \bibinfo{journal}{Physical Review A} \textbf{\bibinfo{volume}{72}},
  \bibinfo{pages}{52331} (\bibinfo{year}{2005}).

\bibitem[{18}]{horodecki1998mse}
\bibinfo{author}{\bibfnamefont{M.}~\bibnamefont{Horodecki}},
  \bibinfo{author}{\bibfnamefont{P.}~\bibnamefont{Horodecki}},
  \bibnamefont{and}
  \bibinfo{author}{\bibfnamefont{R.}~\bibnamefont{Horodecki}},
  \bibinfo{journal}{Physical Review Letters} \textbf{\bibinfo{volume}{80}},
  \bibinfo{pages}{5239} (\bibinfo{year}{1998}).

\bibitem[{19}]{lewenstein2000oew}
\bibinfo{author}{\bibfnamefont{M.}~\bibnamefont{Lewenstein}},
  \bibinfo{author}{\bibfnamefont{B.}~\bibnamefont{Kraus}},
  \bibinfo{author}{\bibfnamefont{J.}~\bibnamefont{Cirac}}, \bibnamefont{and}
  \bibinfo{author}{\bibfnamefont{P.}~\bibnamefont{Horodecki}},
  \bibinfo{journal}{Physical Review A} \textbf{\bibinfo{volume}{62}},
  \bibinfo{pages}{52310} (\bibinfo{year}{2000}).

\bibitem[{20}]{sanpera2001snw}
\bibinfo{author}{\bibfnamefont{A.}~\bibnamefont{Sanpera}},
  \bibinfo{author}{\bibfnamefont{D.}~\bibnamefont{Bru{\ss}}}, \bibnamefont{and}
  \bibinfo{author}{\bibfnamefont{M.}~\bibnamefont{Lewenstein}},
  \bibinfo{journal}{Physical Review A} \textbf{\bibinfo{volume}{63}},
  \bibinfo{pages}{50301} (\bibinfo{year}{2001}).

\bibitem[{21}]{bertlmann2008gew}
\bibinfo{author}{\bibfnamefont{R.}~\bibnamefont{Bertlmann}} \bibnamefont{and}
  \bibinfo{author}{\bibfnamefont{P.}~\bibnamefont{Krammer}},
  \bibinfo{journal}{Physical Review A} \textbf{\bibinfo{volume}{77}},
  \bibinfo{pages}{24303} (\bibinfo{year}{2008}).

\bibitem[{22}]{horodecki1997sca}
\bibinfo{author}{\bibfnamefont{P.}~\bibnamefont{Horodecki}},
  \bibinfo{journal}{Physics Letters A} \textbf{\bibinfo{volume}{232}},
  \bibinfo{pages}{333} (\bibinfo{year}{1997}).

\end{thebibliography}

\end{document}